\documentclass[a4paper,12pt]{article}
\usepackage{latexsym}
\usepackage{amsmath}
\usepackage{amssymb}
\usepackage{bm}
\usepackage{textcomp}
\usepackage{graphicx}
\usepackage[OT1]{fontenc}
\usepackage[english]{babel}

\usepackage{geometry}
\begin{document}

\linespread{1}

\begin{center}
\textbf{\LARGE{A Bound on Equipartition of Energy}}
\newline

\noindent

\noindent

Nicolo' Masi

INFN \& Alma Mater Studiorum - Bologna University,

Via Irnerio 46, 40126 Bologna, Italy

{\it masi@}{\it bo.infn.it}
\end{center}
\begin{center}
\begin{abstract}

\small{\footnotesize{In this article we want to demonstrate that the time-scale constraints for a thermodynamic system imply the new concept of {\it equipartition of energy bound} (EEB) or, more generally, a thermodynamical bound for the {\it partition} of energy. We theorized and discussed the possibility to put an upper limit to the equipartition factor for a fluid of particles. This could be interpreted as a sort of transcription of the entropy bounds from quantum-holographic sector: the EEB number $\pi^{2}/2 = 4.93$, obtained from a comparison between the Margolus-Levitin quantum theorem and the TTT bound for relaxation times by Hod, seems like a special value for the thermodynamics of particle systems.
This bound has been related to the idea of an extremal statistics and independently traced in a statistical mechanics framework, analyzing the mathematical behavior of the distributions which obey to a thermodynamical statistics with a power law greater than the planckian one.}}

\end{abstract}
\end{center}

\begin{center}

\large{\textbf{I. Summary}}
\end{center}

The present paper is organized as follow:
\begin{enumerate}
	\item Sec. II - There are two well known temporal inequalities for a thermodynamic system, one as a function of the average energy of the system (the Margolus-Levitin theorem), the other as a function of temperature (the TTT bound by Hod). If we compare the two formulas, we can easly find an interesting hint: as there are upper limits for the entropy and for the other physical quantities linked to extremal thermodinamic systems (a black hole, for example), so it's legitimate to think that it must exist an upper limit for the statistical mechanics quantity connected to the two energy and temperature relations, i.e. the energy equipartition factor. This is the first fundamental step with which we can obtain the {\it Equipartition of Energy Bound}, that's worth $\pi^{2}/2 = 4.93$, through the equivalence of the two equations. \clearpage
  \item Sec. III - Then we can attack and analyze this result using the statistical mechanics instruments, for the purpose of separately proving its reasonableness. In fact, we also know that the equipartition factor descends from the statistical properties of the fluid, i.e. from the ratio between the energy density and the number density of the system; so it's strongly related to the choice of the statistics we want to use to describe the subnuclear particle ensemble. Therefore, to support our thesis and retrace the EEB value, we have to search for a fluid that obeys to a statistics which is able to mathematically produce the number 4.93. To do this, we have to know how the equipartition factor behave and explore its link with the statitical mechanics integral power law and the features of the fluids with which they are associated. After that, we can probe alternative thermodynamical statistics with a power law greater than the planckian one.
  \item Sec. IV - At first, it seems a pure mathematical study in the statistical mechanics framework, to find the correct representation for such an extremal statistics that causes the EEB (that we will call {\it metaplanckian}) and to understand the complete hierarchy of the equipartition factors. Then it becomes possible to speculate further around the physical nature of this statistics which saturates the equipartition of energy bound. 
  \item Sec. V - Moreover, it could be interesting to investigate the neat mathematical meaning of the numerical factor $\pi^{2}/2 = 4.93$, which is a trascendental special number that also occurs in the comparison between the Margolus-Levitin theorem and the famous Bekenstein bound. We will show that it's even possible to establish another complementary relationship between the TTT bound and the Bekenstein bound.
  The intertwining of these bounds always leads to $\pi^{2}/2$.
\end{enumerate}

\begin{center}
\large{\textbf{II. The Margolus-Levitin theorem and the TTT bound compared: the EEB}}
\end{center}

Let's begin by putting forward the fundamental relations from which our argument starts.

The Margolus-Levitin theorem [18] states that:
\begin{equation}
\Delta\tau_{q-comp}\geq\frac{\pi\hslash}{2\langle E\rangle} \Rightarrow \tau_{q-comp}^{\min}=\frac{\pi\hslash}{2\overline{E}} 
\end{equation}
On the other hand, the TTT (Time Times Temperature) inequality for the relaxation of a thermodynamic system says:
\begin{equation}
\tau_{relax}\geq\frac{1}{\pi T}\frac{\hslash}{k_{B}} \Rightarrow \tau_{relax}^{\min}=\frac{1}{\pi T}\frac{\hslash}{k_{B}}
\end{equation}

The arrows indicate the minimum temporal values, i.e. the limits of each inequality we will use later.

The first represents the minimum time $\tau_{q-comp}^{\min}$ for quantum computation for a system of average energy $\overline{E}$ assigned, i.e. it's the smallest conceivable time interval to switch from a given state to one orthogonal to it. In the inequality the $\left\langle \right\rangle$ brackets stand for the quantum expectation value of energy (above the ground state) and $\Delta\tau_{q-comp}$ represents a generic computational time interval: the formulation reminds an uncertainty principle. In other words, the processing rate of a quantum system cannot exceed $6\cdot 10^{33}$ operations per second per joule. 

Following the Margolus-Levitin theorem, we can say that once the average energy of the system is determined, the minimal evolving time between two distinct states is also determined. A quantum computer with every operation performed as quickly as the minimal evolving time has reached the maximum speed associated to the average energy. 

Margolus calls ``computronium'' a material medium capable of saturating this bound, to be used as the best programmable matter, perhaps a black hole computer [29].

The TTT relation defines the minimum time $\tau_{relax}^{\min}$ required to ensure that the system returns from a certain state to a reference or equilibrium one, as a function of temperature $T$. This bound for the relaxation time was obtained by Hod [2, 3, 30], and it's based on information theory and thermodynamic arguments.
According to quantum theory, a thermodynamic system has at least one perturbation mode whose relaxation time is $\frac{1}{\pi T}\frac{\hslash}{k_{B}}$, or larger. 

This bound is also deeply connected to the black holes world. In fact, in terms of quasinormal black hole frequencies and temperature, we can say there must exist a quasinormal frequency $\omega$ whose imaginary part satisfies $\pi T_{BH} \geq Im(\omega)$, which is obtained by inverting the TTT inequality [1, 4, 5].

They are both {\it evolution times} or {\it response scales} for a generic thermodynamic system.
Therefore, our quantum system that obeys the laws of statistical mechanics shows a temporal resolution, which correspond quantities that describe the fluid in a complementary way, namely energy and temperature.

We don't want to make a treaty on these bounds, but we want to use them to obtain a noteworthy result and emphasize its statistical mechanics nature. Here we only recall two important considerations:

\begin{enumerate}
	\item The constraint by Margolus-Levitin (ML) is a fundamental theorem of modern quantum theory, related to the Mandelstam-Tamm (MT) inequality $\Delta\tau\geq\frac{\pi\hslash}{2\Delta E}$ --- from which it differs for the $\Delta E$ instead of $E$ --- and, of course, to the uncertainty principle. The ML relation is an improved version of the MT one, which in turn is a refinement of the Heisenberg uncertainty principle for the time variable [15, 16, 17]. For further details and a covariant formulation of ML refer to [13].
It should be noted that could exist, even if not yet discovered, quantum estimates more stringent than current ML value, $\mathrm{i}.\mathrm{e}$. a more fundamental {\it temporal computational quantum}.
  \item The thermodynamic TTT bound [6, 7, 8] is saturated by extremal black holes. It's also closely connected to the bound of space discretization discussed by Pesci $l^{*}=1/\pi T$, with which there are strong similarities. This means that it could be linked to the new ideas of general relativity space-time discretization from the holographic theory, antagonist of string theory. It has a universal meaning and it may represent an ultimate physical scale. For further details refer to [9, 10, 11].
There is not yet a statement of which exactly are the usual systems for which it doesn't hold, but we are going to think about extreme black hole-like systems where it is applicable and nearly saturated.
\end{enumerate}

Going now to match these time scales that characterize a thermodynamic system in which quantum properties reign, we try to investigate the implications of the boundary behavior of the system itself. We want to move towards a new original bound to shed some more light on the border, or {\it frontier}, thermodynamic properties --- in analogy with t'Hooft Holographic Principle [19, 20, 21]. We'll see that the result will have an autonomous and independent relevance.

Now we assume that $\tau_{q-comp}^{\min}\geq\tau_{relax}^{\min}$ , i.e. an order relation that says the minimum thermodynamic relaxation time is less than or equal to the minimum quantum computational one: there might be a perturbation mode whose time of relaxation is shorter than that of the fastest flip between quantum states. 

Then, equaling the two temporal equation in (1) and (2), respectively in $\overline{E}$ and $T$, we derive the upper limit --- a dimensionless bound --- for the ratio $\overline{E}/T$ :
\begin{equation}
 \tau_{q-comp}^{\min}\equiv\tau_{relax}^{\min}\Rightarrow\frac{\overline{E}}{k_{B}T}=\frac{\pi^{2}}{2}=4.9348
\end{equation}
The law thus obtained has the form of a quantum equipartition principle of a {\it marginal land} (where these two times can be considered equal, since in general they're not equal), for fluids that satisfy, as mentioned above, the property of extremal blacks holes and ideal computers. 

We have found an {\it Equipartition of Energy Bound} which has to do with physical systems that saturate the two temporal bounds and minimize the response scales. We will hereafter shorten with $EEB$. We will see in Sec. III that ordinary matter can't reach this value. This result will be deepened in Sec. V through a cross-comparison with the Bekenstein bound.

Looking at the formula (3), we can also say that energy is classically a function of the $\pi^{2}\sim 10$ degrees of freedom of the system (if the latter assumed quadratic in an unknown Hamiltonian).

Moreover, the left-hand side of the equation (3) of the equipartition of energy (or, generically, {\it of the energy partition}), can be explicitly written according to the requirements of statistical mechanics [25, 28] and to the previous temporal quanta inequality:
\begin{equation}
\beta\overline{E}=\beta\frac{\int_{0}^{\infty}\epsilon N(\epsilon)d\epsilon}{\int_{0}^{\infty}N(\epsilon)d\epsilon}=f_{eq}\leq 4.93
\end{equation}

\noindent where $\beta$ is the Boltzmann factor $(k_B T)^{-1}$, $\epsilon$ the energy variable, $N(\epsilon)$ is the particle population as a function of energy, determined by fundamental unfamiliar constituents, which obey a statistics to be defined. We call $f_{eq}$ the numerical equipartition factor that we are going to examine. 

Equations (3) and (4) are the heart of the whole discussion.

As is usual in the holographic field, we have an information ``on the border'' that comes indirectly from theory (i.e. by the equality of the time constraints), but that does not clearly show us the nature of the degrees of freedom at stake.

We want to stress that the numerical value 4.93 we obtained comes from a comparison of two fundamental relations, which are independent of the actual calculations of statistical mechanics.
We have also to remember that the EEB is a gravitational-free result, which do not depends on Newton's gravitational constant $G_{N}$, unlike holographic bounds.

\begin{center}
\large{\textbf{III. On the Energy Equipartition Factors}}
\end{center}
Below, we try to sort and outline the basic useful properties of thermodynamic systems made up of non-interacting particles.
First, we remind that the most general form of the equipartition theorem [23] states that, for a physical system with Hamiltonian energy function $H$ and degrees of freedom $x_n$, the following equipartition formula holds in thermal equilibrium for all indices {\it m} and {\it n}:
\begin{equation}
    \Bigl\langle x_{m} \frac{\partial H}{\partial x_{n}} \Bigr\rangle = \delta_{mn} k_{B} T. 
\end{equation}
Here $\delta_{mn}$ is the Kronecker delta, which is equal to one if {\it m = n} and is zero otherwise. The averaging brackets $\left\langle \right\rangle$ is assumed to be an ensemble average over phase space or, under an assumption of ergodicity, a time average of a single system. With this theorem we can derive the equipartition factors for classical, ultrarelativistic or not, particle systems.

The quantum equipartitions factors, for fermions and bosons, are obtained from statistical mechanics ratio between the energy density $\mathcal{E}$ and the number density $n$ of the particle gas. The ratio can be written, in natural units, for both Fermi-Dirac and Bose-Einstein statistics. 
\begin{equation}
f_{eq}=\frac{\mathcal{E}}{n} =\frac{1}{T}\int_{0}^{\infty}\frac{\epsilon^{\alpha}}{e^{\frac{\epsilon}{T}}\pm1}d\epsilon\bigg/\int_{0}^{\infty}\frac{\epsilon^{\alpha-1}}{e^{\frac{\epsilon}{T}}\pm1}d\epsilon 
\end{equation}

\begin{table}[h!]
\begin{center}
\begin{tabular}{c c c c}\hline
    & $f_{eq}$ & $\epsilon^{\alpha}\sim\epsilon N(\epsilon)$ & $w$ \\ \hline
\textit{\textbf{Massive bosons}} & $0.77$ & $\epsilon^{3/2}$ & $2/3$ \\ \hline
\textit{\textbf{Classical particles}} & $1.50$ & $\epsilon^{3/2}$ & $2/3$ \\ \hline
\textit{\textbf{Fermions}} & $1.70$ & $\epsilon^{3/2}$ & $2/3$ \\ \hline
\textit{\textbf{Photons}} & $2.70$ & $\epsilon^{3}$ & $1/3$ \\ \hline
\textit{\textbf{Ultrarelativistic classical particles}} & $3.00$ & $\epsilon^{3}$ & $1/3$ \\ \hline
\textit{\textbf{Ultrarelativistic fermions}} & $3.15$ & $\epsilon^{3}$ & $1/3$ \\ \hline
\end{tabular}
\end{center}
\caption{Equipartition factors for all the combination of known massive/massless fermions/bosons.}
\end{table}

In Table 1, $\epsilon^{\alpha}$ represents the energetic power law that also appears in the numerator of equation (4). You can note the equipartition factors grow together with the energy power, where bosons precede fermions both in $\alpha=3/2$ and $\alpha=3$ domains, and classical particles are in the middle. The values modulation represents the deviation from the theoretical classical Boltzmann behavior. For example, for an ultrarelativistic gas of fermions, like a white dwarf or a neutron star, the equipartition factor is more than double that classical one. This is the higher factor we know. 

As you can see, the value 4.93 should really be an upper limit for this sequence.
This particle equipartition factors hierarchy is also reflected in the analysis of thermodynamic gravitating systems, as non-viscous fluids described in the usual way by a linear EOS $ p=w\rho$.
As shown by Pesci in [12], it is possible to highlight a genuine saturation hierarchy of the noteworthy bound; for example, in terms of constraints to the ratio $M/R$, mass divided by radius, or to the extensivity factor of a thermodynamic system, with the Oppenheim method [27]. 
This is the crux of the argument: we suggest that, increasing $\alpha$, we approach a limit of several physical quantities: if for $w =2/3$ and $\alpha = 3/2$ you are very far from the bounds (non-relativistic matter), you get closer and closer for $w =1/3$ and $\alpha = 3$.

\begin{center}
\large{\textbf{IV. A Possible Origin for the Equipartition of Energy Bound}}
\end{center}
Trying to continue this thermodynamical categorization and reach a possible origin for the equipartition energy bound, you can analyze some extended bosonic statistical laws, that will naively call ``metaplanckian'', in order to bridge the gap that lies between the energy equipartition factor for relativistic fermions, 3.15, and the EEB, that's 4.93. So we are searching for statistical laws able to suggest well-placed values on the thermodynamical hierarchy illustrated in Table 1.

We assume that these extreme fluids have a bosonic nature (or {\it pseudobosonic}, because of their special features), satisfying the Bose-Einstein occupation law $\ (e^{\epsilon/T}-1)^{-1}$, with $k_{B}=1$, though with a modified factor of degeneration $g(\epsilon)$ of the states.

This is reasonable since we are studying something not ordinary, that must be sufficiently stable, lacking any fermionic nature and effects of the Pauli exclusion principle; something that behaves like a scalar or an equally minimal {\it object}, as could be a bosonic condensate. The power law must then show an $\alpha\in \mathrm{N}$ and overcome the planckian dependence. We can study the $\alpha=4$ and $\alpha=5$ cases and explicitly calculate the properties of statistical mechanics analytical functions [22, 24, 25, 26]: energy density, number density and, finally, {\it equipartition factor} (or partition factor, because we don't know the exact meaning of the degrees of freedom at stake).
\begin{equation}
\mathcal{E}=\frac{1}{8\pi^{3}M^{\alpha-3}}\int_{0}^{\infty}\frac{4\pi\epsilon^{\alpha}}{e^{\frac{\epsilon}{T}}-1}d\epsilon=\frac{M^{3-\alpha}Li_{\alpha+1}(1)T^{\alpha+1}\Gamma(\alpha+1)}{2\pi^{2}}=\big\{\frac{12T^{5}\zeta(5)}{\pi^{2}M},\frac{4\pi^{4}T^{6}}{63M^{2}}\big\}
\end{equation}

\begin{equation}
n=\frac{1}{8\pi^{3}M^{\alpha-3}}\int_{0}^{\infty}\frac{4\pi\epsilon^{\alpha-1}}{e^{\frac{\epsilon}{T}}-1}d\epsilon=\frac{M^{3-\alpha}Li_{\alpha}(1)T^{\alpha}\Gamma(\alpha)}{2\pi^{2}}=\big\{\frac{\pi^{2}T^{4}}{30M},\frac{12T^{5}\zeta(5)}{\pi^{2}M^{2}}\big\}
\end{equation}
\begin{equation}
f_{eq}^{B}=\frac{\mathcal{E}}{n} =\frac{1}{T}\int_{0}^{\infty}\frac{\epsilon^{\alpha}}{e^{\frac{\epsilon}{T}}-1}d\epsilon\bigg/\int_{0}^{\infty}\frac{\epsilon^{\alpha-1}}{e^{\frac{\epsilon}{T}}-1}d\epsilon=\{3.83, 4.91\} 
\end{equation}

In curly brackets there are the results for $\alpha=4,5$.
As we said, with this set of formulas we can also reproduce all the values in Table 1 (with {\it plus} for fermions). They are completely general formulas, with which every standard statistical mechanics result can be derived, along with some new outcomes, using $\alpha > 3$.

We have assumed that $\mu=0$, that is a solid condition for known bosons $\gamma$, g, W, Z, both massive and massless; this can also be interpreted as a condition of distance from the degeneration $\mu << T$. The function $Li_{\alpha}(1)$ is the polylogarithm, estimated when the fugacity $z=e^{\mu/T}$ goes to 1, $\zeta(5)$ is the Riemann Zeta function, and $\Gamma(\alpha)$ the Euler Gamma. Degeneration factors $g(\epsilon)$ of the statistics (or densities of states), are not explicitly represented, since their numerical parts cancel each other in the calculation of $f_{eq}$. So they appear only indirectly, as carriers of the $\alpha$-th power law. The $M^{3-\alpha}$ mass factor has been introduced to extend the dimensional validity of classical formulas. You can see that the mass factor vanishes for photons.

The equipartition coefficients thus obtained, 3.83 and 4.91, perfectly fill the gap. The second factor appears almost identical to the critical value previously calculated by comparing the quantum-holographic bounds, with a discrepancy of the order of \%. As mentioned in Sec. II, there might exist a more stringent estimate, capable of lowering 4.93 to 4.91, or some quantum corrections of higher order.
So, using $\alpha=5$ we obtained a proper coefficient which completes the hierarchy of the thermodynamics of particle species, up to the EEB. The coefficient belong to a scalar fluid, which have the Bose-Einstein term at the denominator and a {\it metaplanckian} power law at the numerator.

Therefore, the equipartition of energy bound can be seen as the first {\it attempt to transcription} of the quantum-holographic entropy bounds in statistical mechanics framework: $\pi^{2}/2$ constrains the ratio between the average energy of a system and its temperature and it was also shown that it could be independently derivable from the use of statistical distributions.

It should be noted that the maximum equipartition factor we obtained in (3) can also be read as a prescription on $\alpha$, which implies $\alpha\leq 5$: hence it defines an {\it extremal statistical distribution}. We may affirm that we can't conceive (or nature can't arrange) particle gases with {\it pseudobosonic} statistical laws with a power grater than 5, even if mathematically possible.

To exhaust the subject, let's take a quick look at the behavior of bosonic equipartition factors in general, depending on the fugacity of the thermodynamic system, for $\alpha\geq 3/2$, i.e. for non-relativistic pressure-carriers fluids, for relativistic and metaplanckian ones, up to $\alpha=8$. The choice of $\alpha = 8$ is arbitrary: here we only want to show the behavior of some equipartition factors beyond $\alpha = 5$.

\begin{figure}[htp]
\centering
\includegraphics[height=7cm]{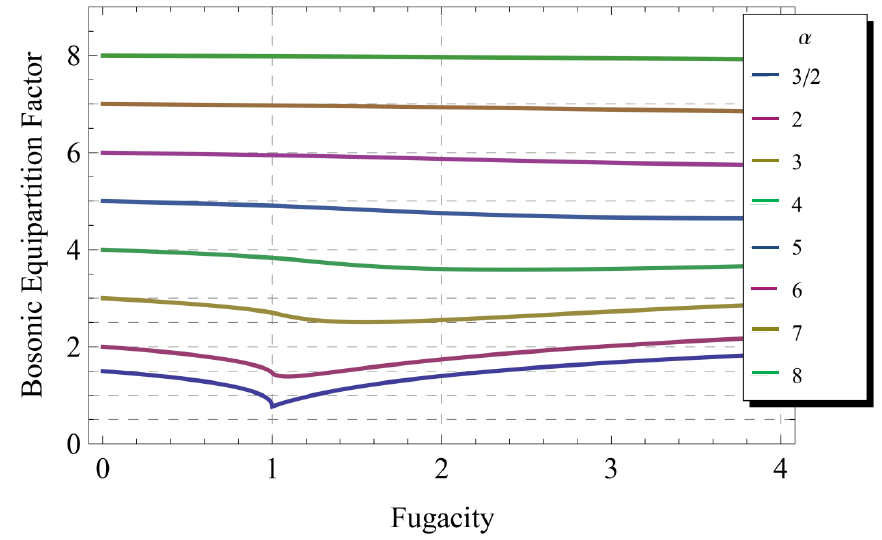}
\caption{Equipartition factor for different bosonic fluids.}
\label{fig:Equip}
\end{figure}

The analytical expression for the bosonic equipartition factor is:
\begin{equation}
f_{eq}^{B}=\displaystyle \alpha\bigg|{\frac{Li_{\alpha+1}(\mathrm{z})}{Li_{\alpha}(\mathrm{z})}}\bigg|
\end{equation}
The equipartition coefficients as a function of fugacity, are rational numbers, characterized by small deviations or cusps of the function around the unitary fugacity. After saturated the equipartition of energy bound, for $\alpha>5\in \mathrm{N}$, the coefficients tend to straight lines, i.e. to natural numbers with step 1, as $ f_{eq}^{B}\rightarrow\alpha$. We may conclude that our spin-statistical world lies in the cusps of the plot, and when the cusps vanish, it becomes impossible to have a fluid with a real relation between energy and temperature, or a meaningful concept of equipartition of energy.\newline
\\
\\

\begin{center}
\large{\textbf{V. Another Clue}}
\end{center}
Another symptom of the need for an EEB is also traceable in [13], where the covariant version of Margolus-Levitin theorem applied to a system that satisfies the Bekenstein bound (UEB) leads to a result which we shall summarize, in natural units, as
\begin{equation}
S_{mat}\leq\frac{\pi^{2}}{2}\#_{op}^{ML} 
\end{equation}
\begin{equation}
\#_{op}^{ML}=\frac{2\Delta\tau_{comp}\langle E\rangle}{\pi\hslash}
\end{equation}

\noindent where {\it mat} stands for matter, $\#_{op}^{ML}$ is the maximum number of quantum computational operations that the system of assigned energy can process in a time interval $\Delta\tau_{comp}=2R$ (in natural units), i.e. that is equal to the diameter of the physical system under analysis. If the time interval corresponds to the ML minimum, we can set $\#_{op}^{ML}$ to 1 to highlight the $\pi^{2}/2$ bound.

Then we can reproduce the same argument, leading to $S_{mat}\leq\pi^{2}/2$, this time comparing the TTT and the Bekenstein bound, in order to complete all possible comparisons between the three bounds. Here we are treating these bounds as a sort of three time-energy (or time-temperature) uncertainty principles with different natures, belonging to three different conceptual areas.

So, if we use $\overline{E} = f_{eq} T$ and $R \leq \tau_{BEK} / 2 $, and substitute them in the UEB, we obtain

\begin{equation}
S_{mat}\leq2\pi R\overline{E}  \Rightarrow  \tau_{BEK}\geq\frac{S_{mat}}{\pi f_{eq} T}
\end{equation}
Stating that $\tau_{BEK}^{\min}\leq\tau_{relax}^{\min}=(\pi T)^{-1}$, we get 
\begin{equation}
S_{mat}\leq f_{eq}\leq\pi^{2}/2
\end{equation}
For the second inequality we have used the EEB.
Thus, even linking the ML bound with the UEB and the TTT bound with the UEB, we can extract entropy bounds defined by a special number.
Recovering the time inequality in Sec. II, we can write the full time inequality
\begin{equation}
\tau_{q-comp}^{\min}\geq\tau_{relax}^{\min}\geq\tau_{BEK}^{\min} 
\end{equation}
This is a hierarchy of temporal quanta from the three relations, and it gives rise to the trascendental limit $\pi^{2}/2$ for each comparison, both for entropy and equipartion factor. $\tau_{BEK}$ seems to be the more fundamental one. 

Regardless of the chain of temporal inequalities, it is a posteriori obvious that 4.93 is a maximum and not a minimum, both for the entropy and the equipartition factor. In fact, on the one hand $f_{eq}$ grows from the value 0.77 to 3.15, while the entropy has its minimum at zero. Therefore, this can also be a criterion to get back to the relationship (15) between the minimum times of the three relations and it is an {\it a posteriori} evidence of the assumption that led to the EEB statement.
This apparent coincidence is worthy of further study.

\begin{center}
\large{\textbf{VI. Conclusions}}
\end{center}
We first demonstrated that from the comparison of time uncertainty relations for quantum thermodynamic systems, assuming an order relation between temporal quanta, you can get a kind of energy-temperature uncertainty relation. The latter translates into an {\it equipartition of energy bound}, i.e. into a maximum physical value for the proportionality coefficient between everage energy and temperature. 

The numerical trascendental EEB factor $\pi^{2}/2 = 4.93$, obtained from the ``symbiosis'' of the Margolus- Levitin quantum theorem with the TTT relaxation bound by Hod, seems to be the largest value physically, but not mathematically, accessible of an increasing sequence of equipartition factors derived from the study of particle fluids. It represents a constraint on how to organize the energy of a system, according to the temperature and the degrees of freedom.

Secondly, the special value 4.93, that may have a quantum-holographic origin, finds an intriguing match within the extreme fluids that obey to bosonic {\it metaplanckian} laws. In fact, the EEB value can be independently traced in a statistical mechanics framework. Using statistical mechanics formulas and $\alpha-th$ power laws greater than the planckian one, we found an equipartition factor ($\alpha=5$) which almost saturates the EEB.

This new bound seems to have something to do with the other famous entropy bounds, as deduced from the application of Margolus-Levitin theorem to a Bekenstein system and from the later comparison between the TTT bound and the Bekenstein bound.

\begin{center}
\large{\textbf{References}}
\end{center}

[1] S. Hod, {\it Phys. Rev. D}75 (2007) 064013

[2] S. Hod, {\it Gen. Rel. Grav.} Vol. 41, No. 10, 2295-2299, Springer

[3] S. Hod, {\it Class. Quant. Grav}. 24, 4235 (2007) arXiv:0705.2306

[4] S. Hod, {\it Phys. Rev. D}78, 084035 (2008) 9 arXiv:0811.3806

[5] S. Hod, {\it Phys. Lett. B}666,483 (2008), arXiv:0810.5419

[6] K. Ropotenko, arXiv:0705.3625

[7] A. Gruzinov, arXiv:0705.1725

[8] A. Lopez-Ortega, {\it Int. Jour. Mod. Phys. D}, Vol. 19, Issue 12 (2010)

[9] A. Pesci, arXiv:0708.3729v2

[10] A. Pesci, arXiv:0803.2642v2

[11] A. Pesci, arXiv:0807.0300v2

[12] A. Pesci, arXiv:0611103v2

[13] Q. Cao, Y.-X. Chen, J.-L. Li, arXiv:0805.4250v1

[14] M. Andrecut, M. K. Ali, J. Phys. A: Math. Gen. 37 (2004) L157-L160

[15] U. Yurtsever, arXiv:0912.4505v3

[16] B. Zielinski, M. Zych, arXiv:quant-ph/0603076v4

[17] P. J. Jones, P. Kok, arXiv: 20034870v1

[18] N. Margolus, L. B. Levitin, {\it Physica D} (1998) Vol. 120, 188-195

[19] P. A. Zizzi, {\it Entropy} 2000, 2, 39-69

[20] B. Fauser {\it et al}., {\it Quantum Gravity}, Birkhauser Verlag, 2007

[21] R. Bousso, {\it Rev. Mod. Phys.} 74, 825-874 (2002) 

[22] J. F. Annett, Oxford University Press 2003

[23] D. A. McQuarrie, {\it Statistical Mechanics}, University Science Books, 2000

[24] K. Huang, {\it Statistical Mechanics}, John Wiley and Sons, 1987

[25] L. D. Landau, E. M. Lifshitz, {\it Statistical Physics}, Pergamon Press, 1980

[26] B. H. Lavenda, A {\it New Perspective of Thermodynamics}, Springer 2010

[27] S. R. Addison, J. E. Gray, {\it J Phys}. A: {\it Math. Gen}. 347733, 2001

[28] S. Lloyd, arXiv:quant-ph/9908043v3

[29] S. Lloyd, Y. Jack Ng, {\it Black Hole Computers}, Scientific American (2004)

[30] S. Hod, {\it Phys. Rev. D} 75, 064013 (2007)

\end{document}